\documentclass [12pt]{article}
\usepackage{graphicx,amssymb,amsfonts,latexsym,amsmath,amsthm,times}
\usepackage{epsfig}
\usepackage{fancyhdr}
\usepackage{color}
\usepackage[english]{babel}
\numberwithin{equation}{section}

\begin{document}

\footnotesize {\flushleft \mbox{\bf \textit{Math. Model. Nat.
Phenom.}}}
 \\
\mbox{\textit{{\bf Vol. 10, No. ?, 2015, pp. ??-??}}}

\thispagestyle{plain}

\vspace*{2cm} \normalsize \centerline{\Large \bf Magnetic Flux Leakage Method: Large-Scale Approximation}

\vspace*{1cm}

\centerline{\bf Anastasiya V. Pimenova$^{a,}$\footnote{Corresponding
author. E-mail: Anastasiya.Pimenova@gmail.com},
                Denis S. Goldobin$^{a,b,c}$,
                Jeremy Levesley$^b$,}
\centerline{\bf Andrey O. Ivantsov$^a$,
                Peter Elkington$^d$,
                Mark Bacciarelli$^d$}

\vspace*{0.5cm}

\centerline{$^a$ Institute of Continuous Media Mechanics, UB RAS, Perm 614013, Russia}

\centerline{$^b$ Department of Mathematics, University of Leicester, Leicester LE1~7RH, UK}

\centerline{$^c$ Department of Theoretical Physics, Perm State University, Perm 614990, Russia}

\centerline{$^d$ Weatherford, East Leake, Loughborough LE12~6JX, UK}


\vspace*{1cm}

\noindent {\bf Abstract.}
We consider the application of the magnetic flux leakage (MFL) method to the detection of defects in ferromagnetic (steel) tubulars. The problem setup corresponds to the cases where the distance from the casing and the point where the magnetic field is measured is small compared to the curvature radius of the undamaged casing and the scale of inhomogeneity of the magnetic field in the defect-free case. Mathematically this corresponds to the planar ferromagnetic layer in a uniform magnetic field oriented along this layer. Defects in the layer surface result in a strong deformation of the magnetic field, which provides opportunities for the reconstruction of the surface profile from measurements of the magnetic field. We deal with large-scale defects whose depth is small compared to their longitudinal sizes---these being typical of corrosive damage. Within the framework of large-scale approximation, analytical relations between the casing thickness profile and the measured magnetic field can be derived.

\vspace*{0.5cm}




\vspace*{1cm}


\section{Introduction}
The magnetic flux leakage (MFL) method is a powerful tool for non-distructive inspection of the integrity of ferromagnetic casings~\cite{Ida-Lord-1983,Al-Naemi-Hall-Moses-2006,Li-Wilson-Tian-2007,Miller-Sander-2008,Sharar-Cuthill-Edwards-2008,Saha-etal-2010,Katoh-Nishio-Yamaguchi-2004} or, more generally, for determining the shape of ferromagnetic objects. The basic idea of the method is to reconstruct the shape features of the inspected ferromagnetic object from the deformation of the magnetic field. The conventional way of applying the MFL method is (i)~recognition of typical patterns in the magnetic field measurement data, (ii)~identification of typical defects of the casing corresponding to these patterns, and (iii)~evaluation of geometrical parameters of the defects identified from the patterns. Such an approach faces obvious problems when one deals with complex-shape defects such as metal losses due to corrosion. Although the problem of the reconstruction of an arbitrary object shape from measurements of the magnetic field (all three components of the $\vec{H}$-field should be measured) on some surface in space near the object is mathematically well-posed and resolvable (numerically), in practice one encounters issues which make this approach generally inapplicable (which is the reason for using the conventional practice outlined above). However, specifically in the case where complex-shape defects are most important---in the case of corrosion damage---there is an opportunity for reconstruction of the arbitrary casing thickness profile based on an analytical approach. This analytical approach is possible owing to the model reduction---the large-scale approximation, which assumes the defect depth and the casing thickness to be small compared to the defect width, as it is typical for corrosion metal losses. The corrosion loss of metal can be non-large-scale near a weld, where corrosion rapidly advances along the contact interface. Otherwise, the large-scale approximation is reasonable.

\begin{figure}[!t]
\center{
\includegraphics[width=0.35\textwidth]%
 {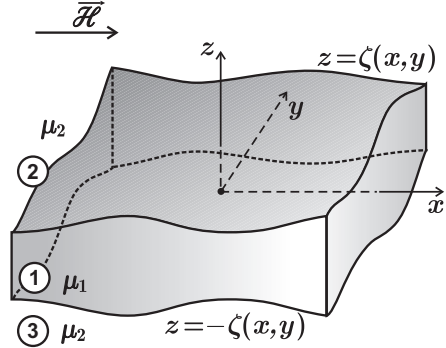}
}

  \caption{
Mirror-symmetric ferromagnetic layer of non-uniform thickness in the magnetic field $\mathcal{H}$ and the coordinate frame}
  \label{sketch}
\end{figure}

In this paper we consider the application of the MFL method to the detection of defects in ferromagnetic (e.g.\ steel) tubulars including wellbore casings. Our treatment is focused on the accurate reconstruction of casing thickness profiles, in contrast to the conventional approach which is the recognition of magnetic field patterns corresponding to catalogued typical defects by means of neural networks or similar data analysis tools. The problem setup we use corresponds to measurements made with modern devices designed for the MFL inspection of wellbore casings (e.g., see~\cite{Sharar-Cuthill-Edwards-2008}). In this setup the distance from the casing and the point where the magnetic field is measured is small compared to the curvature radius of the undamaged casing and the scale of inhomogeneity of the magnetic field in the defect-free case. Mathematically, this corresponds to the planar ferromagnetic layer in a uniform magnetic field oriented along this layer. Defects of the layer surface result in a strong deformation of the magnetic field, which provides opportunities for the reconstruction of the surface profile from the measurements of the magnetic field.

\setcounter{equation}{0}
\section{Analytical theory: Large-scale approximation}
We restrict the analytical treatment to the case of large-scale defects, i.e.\ defect length and width is large relative to the depth or the casing (metal layer) thickness. For this case we show that the magnetic field is sensitive to the casing thickness profile, but not to the inner and outer surface profiles independently. Hence, as a starting point, one can address the problem of a symmetric layer, one surface of which is a mirror image of the other.

\subsection{Mirror-symmetric layer}
\subsubsection*{Mathematical description of the problem}
Let us consider the ferromagnetic layer confined between two surfaces $z=\zeta(x,y)$ and $z=-\zeta(x,y)$, where the $(x,y)$-plane is the middle plane of the layer and the $z$-axis is orthogonal to it. The uniform external magnetic field $\mathcal{H}$ is applied along the $x$-axis. The system is sketched in Figure~\ref{sketch}.

We adopt the following assumptions for the problem:
\begin{enumerate}
\item
The layer geometry and fields possess the symmetry property
($z\rightarrow-z$);
\item
The linear magnetisation law for both the ferromagnetic material and the material around it is given by: $\vec{B}_j=\mu_j\vec{H}_j$, $j=1,2$;
\item
The magnetic permeability of the surrounding material is small
compared to that of the ferromagnet,
 $\displaystyle \frac{\mu_1}{\mu_2}\gg1$;
\item
Surface defects are large-scale, which means the typical
longitudinal size of defects $L\gg\zeta$ and, therefore,
$|\nabla_2\zeta|\ll1$.
\end{enumerate}
(In the following subsections we will extend our consideration beyond restrictions (1) and (2).)
Henceforth, the subscripts of fields and parameters indicate the corresponding domain (1: ferromagnet, 2: upper outer area, 3: lower outer area); for the gradient and Laplace operators, $\nabla_2$ and $\Delta_2$, the index ``2'' indicates the two-dimensional versions of them calculated with respect to $x$ and $y$ coordinates only. The ranges of parameter values of practical interest are presented in Table~\ref{tab1} and are consistent with the assumptions made.

According to Maxwell's equations, we have the following equation system
$$
\left\{
\begin{aligned}
\nabla\times\vec{H}_j&=0\,,\\
\nabla\cdot\vec{B}_j&=0\,,
\end{aligned}
\right.
$$
with boundary conditions
$$
\vec{H}_{1\tau}=\vec{H}_{2\tau}\;,\qquad
B_{1n}=B_{2n}
$$
for the normal to the surface (subscript ``$n$'') and tangential (``$\tau$'') components of magnetic field, respectively. When the curl of a vector field is zero in a simply-connected domain, one can introduce the scalar potential for this field within this domain. We introduce the scalar potential $\Phi$, $\vec{H}=-\nabla\Phi$, which obeys the equation
\begin{equation}
\Delta\Phi_j=0\,,
\label{eq101}
\end{equation}
while the boundary conditions read
\begin{eqnarray}
&\displaystyle
\Phi_1=\Phi_2\;,
\label{eq102}
\\[5pt]
&\displaystyle
\mu_1\frac{\partial\Phi_1}{\partial n}=
\mu_2\frac{\partial\Phi_2}{\partial n}\;.
\label{eq103}
\end{eqnarray}

\begin{table}[!t]
\caption{Reference values of real system parameters}
\begin{center}
\begin{tabular}{lc}
\hline
\\[-10pt]
 ratio of magnetic permeabilities  & $\mu_1/\mu_2$ : \ $100-1000$\\[5pt]
 longitudinal scale of defects & $L$ : \ $(10-20)\zeta$\\[5pt]
 location of magnetic sensors & $z$ : \ $(2-8)\zeta$\\[5pt]
\hline
\end{tabular}
\end{center}
\label{tab1}
\end{table}

Since the magnetic permeability of the ferromagnet is considerably larger than that of the surrounding material, to the leading order of approximation the flux of the magnetic field does not go outside the boundaries of the ferromagnet, i.e.
\begin{equation}
\nabla_2\left(\int_{-\zeta}^\zeta
\left(-\frac{\partial\Phi_1}{\partial x}\right)dz
\right)=0\,.
\label{eq104}
\end{equation}
In this case the normal derivatives of $\Phi$ on the boundary are equal to zero, and Eq.\,(\ref{eq103}) takes the form
\begin{equation}
\frac{\partial\Phi_1}{\partial n}=0\,.
\end{equation}

For infinitely large scale inhomogeneities the magnetic $H$-field within the layer is the same as for the defect-free planar layer, $\vec{H}_1=\vec{\mathcal{H}}$. Hence, we can look for the correction to the uniform field $\vec{\mathcal{H}}$. One can write down the Taylor series for $\Phi_1$ with respect to $z$
$$
\Phi_1(x,y,z)=-\mathcal{H}x+\Phi^{(0)}_1(x,y)
 +\Phi^{(2)}_1(x,y)\frac{z^2}{2!} +\Phi^{(4)}_1(x,y)\frac{z^4}{4!}
 +\dots\;
$$
(only even powers of $z$ are present due to the symmetry $z\to-z$). Substituting this series into Eq.\,(\ref{eq101}) and renaming $\Phi^{(0)}_1(x,y)$ as $F(x,y)$, one finds
\begin{equation}
\Phi_1(x,y,z)=-\mathcal{H}x+F(x,y)
 -\Delta_2 F(x,y)\frac{z^2}{2!} +\Delta_2^2 F(x,y)\frac{z^4}{4!}
 -\dots\;.
\label{eq106}
\end{equation}
Notice, here the $z^{2n}$-term is of the order of magnitude of $F(\zeta/L)^{2n}$ and thus only first several terms can be important for the large-scale case. Since $F$ vanishes for infinitely large scale of defects, it should be small compared to the leading term of $\Phi_1$ for finite large scale $L$ by continuity, i.e.\
$$
\big|\nabla_2 F\big|\ll\mathcal{H}\;.
$$

On the surface $z=\zeta(x,y)$, Eq.\,(\ref{eq102}) yields
\begin{equation}
\Phi_2(z=\zeta)=\Phi_1=-\mathcal{H}x+F-\frac{1}{2}\Delta_2 F\zeta^2
 +\mathcal{O}\left(F\frac{\zeta^4}{L^4}\right)\;.
\label{eq107}
\end{equation}

\subsection{Two-dimensional case}
Let us consider the two-dimensional problem of a ferromagnetic layer uniform in the $y$-direction. One can see that for the two-dimensional case the integral in Eq.\,(\ref{eq104}) can be found as
\begin{equation}
 \int_{-\zeta}^\zeta\left(-\frac{\partial\Phi_1}{\partial x}\right)dz=const=2\zeta_0\mathcal{H},
 \label{eq108}
\end{equation}
where $\zeta_0$ is the $z$-coordinate of the undamaged surface. This significantly simplifies the task and makes it possible to solve the problem analytically. Substituting expression~(\ref{eq106}) into the latter equation, one can see that
\begin{equation}
 \frac{\partial F}{\partial x} = \mathcal{H}(1-\frac{\zeta}{\zeta_0})+\mathcal{O}\left(\mathcal{H}\frac{\zeta^3}{L^2}\right).
 \label{eq109}
\end{equation}
$\Phi_2$ can be expanded into a series near the surface $\zeta$;
\begin{equation}
\Phi_2(\zeta)=\Phi_2(h)
 +\left.\frac{\partial\Phi_2}{\partial z}\right|_{z=h}(\zeta-h)
 +\frac{1}{2}\left.\frac{\partial^2\Phi_2}{\partial z^2}\right|_{z=h}
 (\zeta-h)^2
 +\dots\;.
 \label{eq110}
\end{equation}
To calculate $(\partial^2\Phi_2/\partial z^2)$ one can
employ Eq.\,(\ref{eq101}),
\begin{equation}
\left.\frac{\partial^2\Phi_2}{\partial z^2}\right|_{z=h}
 =-\left.\frac{\partial^2\Phi_2}{\partial x^2}\right|_{z=h}.
 \label{eq111}
\end{equation}
Hence, Eq.\,(\ref{eq110}) can be rewritten in the form
\begin{equation}
\Phi_2(h)=-\mathcal{H}x+F-\frac{1}{2}\frac{\partial^2F}{\partial x^2}\zeta^2+\left.\frac{\partial\Phi_2}{\partial z}\right|_{z=h}(h-\zeta)
-\frac{1}{2}\left.\frac{\partial^2 \Phi_2}{\partial x^2}\right|_{z=h}(h-\zeta)^2+\mathcal{O}\left(F\frac{(h-\zeta)^3}{L^3}\right).
\label{eq112}
\end{equation}
Substituting Eq.\,(\ref{eq109}) and differentiating the last equation with respect to $x$, one can evaluate the $x$-component of the magnetic $H$-field measured at the height $h$ above the layer;
\begin{equation}
 \left.H_x\right|_{z=h}=\mathcal{H}\frac{\zeta_0}{\zeta}
 +\frac{\partial}{\partial x}\left(\left.\frac{\partial\Phi_2}{\partial z}\right|_{z=h}(\zeta-h)\right)
 +\mathcal{O}\left(\mathcal{H}\frac{\zeta^2}{L^2}\right).
\label{eq113}
\end{equation}
Let us seek a series expansion for $\zeta=\zeta_1+\zeta_2+\dots$\,, where each term of the series is small compared to the previous one. To the leading order of accuracy the last equation yields
\begin{equation}
\zeta_1=\zeta_0\frac{\mathcal{H}}{\left.H_x\right|_{z=h}}.
\label{eq114}
\end{equation}
Substituting (\ref{eq114}) into Eq.\,(\ref{eq113}) and considering $\vec{H}=-\nabla\Phi$, one can obtain
\begin{equation}
\zeta_2=-\frac{\mathcal{H}\zeta_0^2}{\left.H_x^2\right|_{z=h}}
\frac{\partial}{\partial x} \left(\Big(\frac{\mathcal{H}}{\left.H_x\right|_{z=h}}-1\Big)H_z\right).
\label{eq115}
\end{equation}
Finally, to the terms of order $\left(\mathcal{H}\zeta_0^2/L^2\right)$
\begin{equation}
\zeta\approx\zeta_0\frac{\mathcal{H}}{\left.H_x\right|_{z=h}}
 \left[1-\frac{\zeta_0}{\left.H_x\right|_{z=h}}
 \frac{\partial}{\partial x}
 \left(H_z\left(\frac{\mathcal{H}}{\left.H_x\right|_{z=h}}-1\right)\right)\right].
\label{eq116}
\end{equation}

\subsection{Asymmetric ferromagnetic layer}
In this subsection we argue that for large-scale defects the case of an asymmetric layer is equivalent to the case of mirror-symmetric layer. Let us consider the layer confined between surfaces $z_2=\zeta_2(x,y)$ and $z_3=-\zeta_3(x,y)$ with $\zeta_2\ne\zeta_3$. It is convenient to introduce the middle surface $z_m=\zeta_m(x,y)$.
$$
\zeta_m=\frac{\zeta_2-\zeta_3}{2}\,,
$$
and use the coordinate frame
$$
\widetilde{x}=x\,,
\qquad
\widetilde{z}=z-\zeta_m(x)\,.
$$
Then
$$
\frac{\partial}{\partial x}
 =\frac{\partial}{\partial\widetilde{z}}-\frac{\partial\zeta_m}{\partial\widetilde{x}}\frac{\partial}{\partial\widetilde{z}}
\qquad\mbox{ and }\qquad
\frac{\partial}{\partial z}
 =\frac{\partial}{\partial\widetilde{z}}\;.
$$

In the new coordinates Eq.\,(\ref{eq101}) takes the form
\begin{equation}
\Delta\Phi_i=\widetilde{\Delta}\Phi_i-\left(\frac{\partial^2\zeta_m}{\partial\widetilde{x}^2}
 +2\frac{\partial\zeta_m}{\partial\widetilde{x}}\frac{\partial}{\partial\widetilde{x}}\right)
 +\left(\frac{\partial{\zeta_m}}{\partial\widetilde{x}}\right)^2
 \frac{\partial^2\Phi_i}{\partial\widetilde{z}^2}\,.
\label{eq117}
\end{equation}
For large-scale defects the additional terms of Eq.\,(\ref{eq117}) affects Eqs.\,(\ref{eq104}) and (\ref{eq110}) only in high-order
terms. Hence, Eq.\,(\ref{eq113}) is not affected by asymmetry.

\subsection{Nonlinear magnetisation law}
\begin{figure}[!t]
\center{
\includegraphics[width=0.30\textwidth]%
 {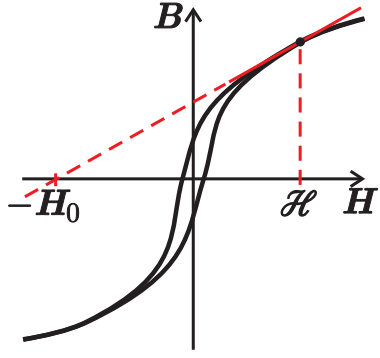}
}

  \caption{
Schematic dependence of the magnetic field $B$ on the magnetic $H$-field in ferromagnetic material (black solid curves) and its linearization in the working parameter range (red line)}
  \label{magnetisation}
\end{figure}

Ferromagnetic materials are characterized by hysteresis and non-linearity of the magnetisation. Hysteresis is to be avoided in inspection applications as it leads to a loss of uniqueness in the solution of the profile-reconstruction problem. For this reason, strong constant magnets are used in practice, and the system operates under conditions close to magnetic saturation (see Figure~\ref{magnetisation}). Close to saturation hysteresis becomes insignificant whereas the non-linearity becomes pronounced and influences applications of the MFL method~\cite{Katoh-Nishio-Yamaguchi-2004}. Nonetheless, non-linear-magnetisation problems can be treated analytically for large-scale defects as well, because the magnetic $H$-field within the ferromagnetic layer deviates slightly from $\vec{\mathcal{H}}$. For brevity, we consider the case of a mirror-symmetric layer in this section.

As $\nabla\times\vec{H}=0$, we still can use substitution
$\vec{H}=-\nabla\Phi$. For $\vec{B}=\mu(H)\vec{H}$, the equation
$\nabla\cdot\vec{B}=0$ takes the form
$$
\nabla\cdot\vec{H}
 +\frac{1}{\mu H}\frac{\partial\mu}{\partial H}
 \vec{H}\cdot\nabla\frac{H^2}{2}=0\;.
$$
Substituting the series
$$
\Phi_1(x,z)=-\mathcal{H}x+\Phi^{(0)}_1(x)
 +\Phi^{(2)}_1(x)\frac{z^2}{2!} +\Phi^{(4)}_1(x)\frac{z^4}{4!}
 +\dots\;
$$
into the latter equation and collecting $z$-free terms, one finds
$$
\frac{\partial^2\Phi_1^{(0)}}{x^2}+\Phi_1^{(2)}=
 -\frac{1}{\mu_1H}\frac{\partial\mu_1}{\partial H}
 \mathcal{H}\frac{\partial}{\partial x}\frac{H^2}{2}+\dots\;.
$$
With
 $H^2=(\mathcal{H}-(\partial/\partial x)\Phi_1^{(0)}+\dots)^2=
 \mathcal{H}^2-2\mathcal{H}(\partial/\partial{x})\Phi_1^{(0)}+\dots$;
to the leading order of accuracy, one can obtain
\begin{align}
 \Phi_1^{(2)}&
 =-\left(1+\beta\right)\frac{\partial^2\Phi_1^{(0)}}{\partial x^2}+\dots\;,
\label{eq118}
\\[5pt]
 \beta&\equiv
 -\frac{\mathcal{H}}{\mu_1(\mathcal{H})}
  \left.\frac{\partial\mu_1}{\partial H}\right|_{H=\mathcal{H}}\;.
\label{eq119}
\end{align}

One can show that $0<\beta<1$. Indeed, let us write $B=\alpha(H+H_0)$ for $H$ next to $\mathcal{H}$ (see Figure~\ref{magnetisation}). Then $\beta=H_0/(\mathcal{H}+H_0)$ with positive $H_0$; therefore, $\beta\in(0,1)$. On the hysteresis loop, which is out of the scope of our study, $\beta$ can be beyond this range.

Eq.\,(\ref{eq102}) holds valid for nonlinear magnetisation and yields a modified version of Eq.\,(\ref{eq106}):
\begin{equation}
\Phi_2(z=\zeta)=\Phi_1=-\mathcal{H}x+\Phi_1^{(0)}
 -\frac{1}{2}\left(1+\beta\right)\frac{\partial^2\Phi_1^{(0)}}{\partial x^2}\zeta^2
 +\dots\,.
\label{eq120}
\end{equation}

Since $0<\beta<1$ for real ferromagnetic materials, the change of a coefficient ahead of the last term does not change its order. As was demonstrated for the case of a mirror-symmetric layer, any terms of this order do not affect the solution to the concerned accuracy. Therefore the nonlinearity of magnetisation does not affect the leading order of the equations we suggest for the profile reconstruction procedure for large-scale defects.

\subsection{Three-dimensional ferromagnetic layer}
For the case of the three-dimensional layer Eq.\,(\ref{eq104}) can be solved only up to the gradient of an arbitrary harmonic function of $x$ and $y$, which is not very helpful for our purposes; the problem requires a somewhat different approach compared to the two-dimensional case.

Let us denote the magnetic potential at the height $\zeta_0$ as $\Phi_2^{(0)}$. One can write down the Taylor series for $\Phi_2$ at the height $\zeta$;
\begin{equation}
\left.\Phi_2\right|_{z=\zeta}=\Phi_2^{(0)}+\left.\frac{\partial\Phi_2}{\partial z}\right|_{z=\zeta_0}(\zeta-\zeta_0)+\frac{1}{2}\left.\frac{\partial^2\Phi_2}{\partial z^2}\right|_{z=\zeta_0}(\zeta-\zeta_0)^2+\dots\,.
\label{eq121}
\end{equation}

Taking into account the  boundary condition $\Phi_1=\Phi_2$, one can equate Eq.\,(\ref{eq106}) with the latter equation and evaluate the leading-order term of the magnetic potential within the layer:
\begin{equation}
\Phi_1(x,y,z)=\left.\Phi_2\right|_{z=\zeta_0}
 +\left.\frac{\partial\Phi_2}{\partial z}\right|_{z=\zeta_0}(\zeta-\zeta_0)
 +\mathcal{O}\Big(\mathcal{H}\frac{\zeta^2}{L}\Big).
\end{equation}

Substituting this series into Eq.\,(\ref{eq104}), one finds
\begin{equation}
\nabla_2\cdot(\zeta\vec{H}
 +\zeta(\zeta-\zeta_0)\nabla_2H_z+\zeta\nabla_2\zeta H_z)=0\,,
\label{eq123}
\end{equation}
where $\vec{H}=-\nabla\Phi_2|_{z=\zeta0}$ is the magnetic field at the height $\zeta_0$ and $H_z$ is its $z$-component. To solve this equation numerically it is convenient to use the exponential representation of $\zeta=\zeta_0e^{-\sigma^{(0)}-\sigma^{(1)}-\dots}$.
To the first two orders Eq.\,(\ref{eq123}) takes the form
\begin{equation}
-\zeta\nabla_2\sigma^{(0)}\cdot\vec{H}-\zeta\nabla_2\sigma^{(1)}\cdot\vec{H}
 +\zeta\nabla_2\cdot\vec{H}+\nabla_2[\zeta(\zeta-\zeta_0)\nabla_2H_z
 +\zeta\nabla_2\zeta H_z]=0\,.
\label{eq124}
\end{equation}
Hence,
\begin{equation}
\nabla_2\sigma^{(0)}\cdot\vec{H}=\nabla_2\cdot\vec{H},
\label{eq125}
\end{equation}
\begin{equation}
\begin{array}{r}
\nabla_2\sigma^{(1)}\cdot\vec{H}
 =\zeta_0\big[(1-3e^{-\sigma{(0)}})\nabla_2\sigma^{(0)}\cdot\nabla_2H_z
 -(1-e^{-\sigma^{(0)}})\Delta_2H_z\qquad\qquad
 \\[7pt]
 {}+2e^{-\sigma^{(0)}}(\nabla_2\sigma^{(0)})^2H_z
 -e^{-\sigma^{(0)}}\Delta_2\sigma^{(0)}H_z\big]\,.
\end{array}
\label{eq126}
\end{equation}
Notice, $\sigma^{(0)}$ is not necessarily small.

With Eqs.\,(\ref{eq125}) and (\ref{eq126}) one can calculate the layer thickness profile $2\zeta=2\zeta_0\exp(-\sigma^{(0)}-\sigma^{(1)}-\dots)$ from the magnetic field $\vec{H}$ (or some of its components) measured at non-large elevation above the layer.

\setcounter{equation}{0}
\section{Application of the analytical technique}
\subsection{Validation of applicability of the analytical technique with results of numerical simulation for two-dimensional case}
In order to validate the applicability of the analytical results derived, we have considered the model case of a ferromagnetic layer of $\mu_2/\mu_1=100$ with profile $\zeta=\zeta_0+a\cos kx$ with $k=2\pi$, $\zeta_0=0.1$, $a=0.01$. The magnetic field for this case was calculated both

\noindent
$\bullet$~with direct numerical simulation, employing a finite volume method and the mesh size $dx=dz=0.01$, and

\noindent
$\bullet$~analytically in Fourier space within the framework of the linear-in-defect approximation.

\begin{figure}[!t]
\center{
{\sf (a)}\hspace{-15pt}
\includegraphics[width=0.65\textwidth]%
  {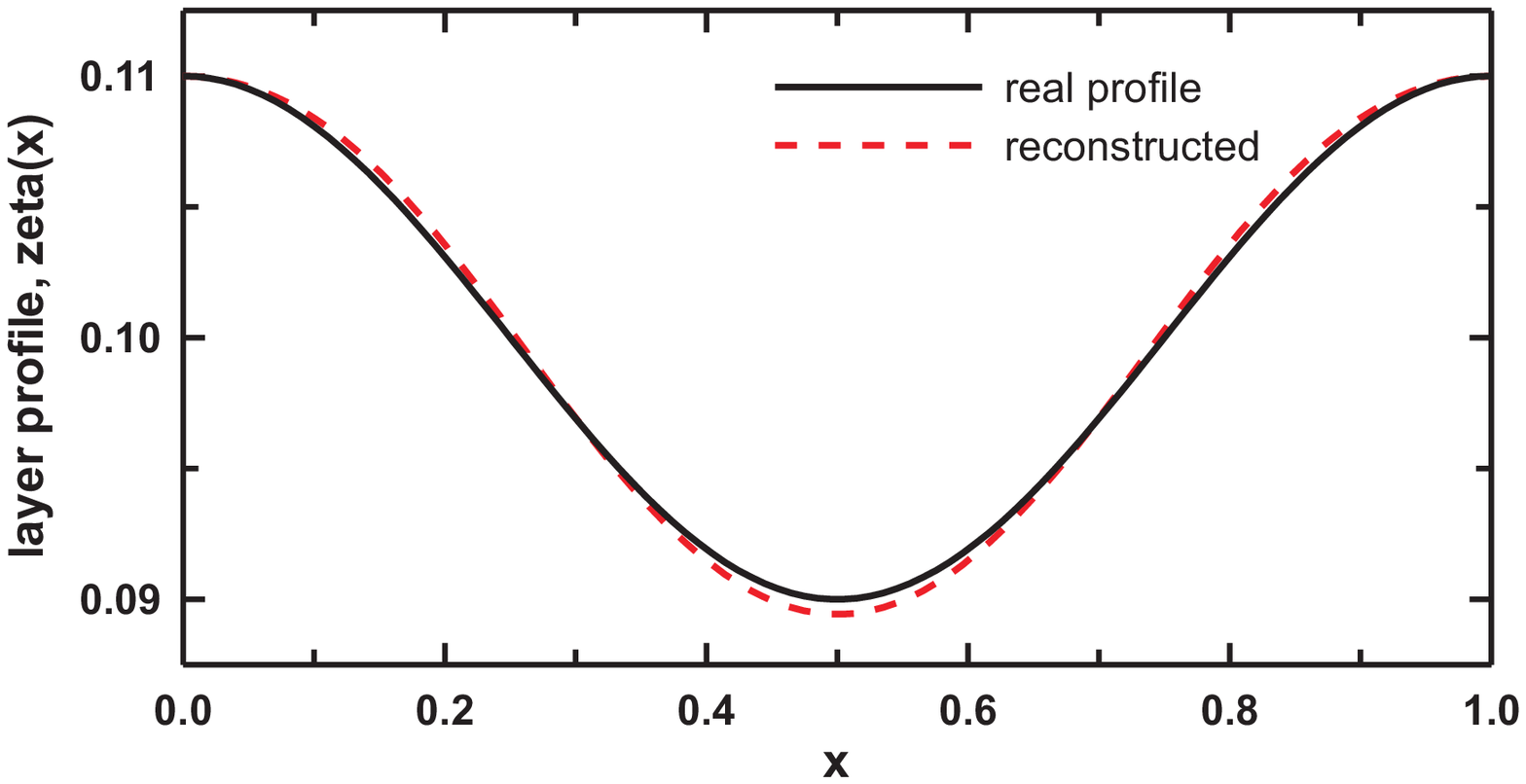}
\\[25pt]
{\sf (b)}\hspace{-15pt}
\includegraphics[width=0.65\textwidth]%
  {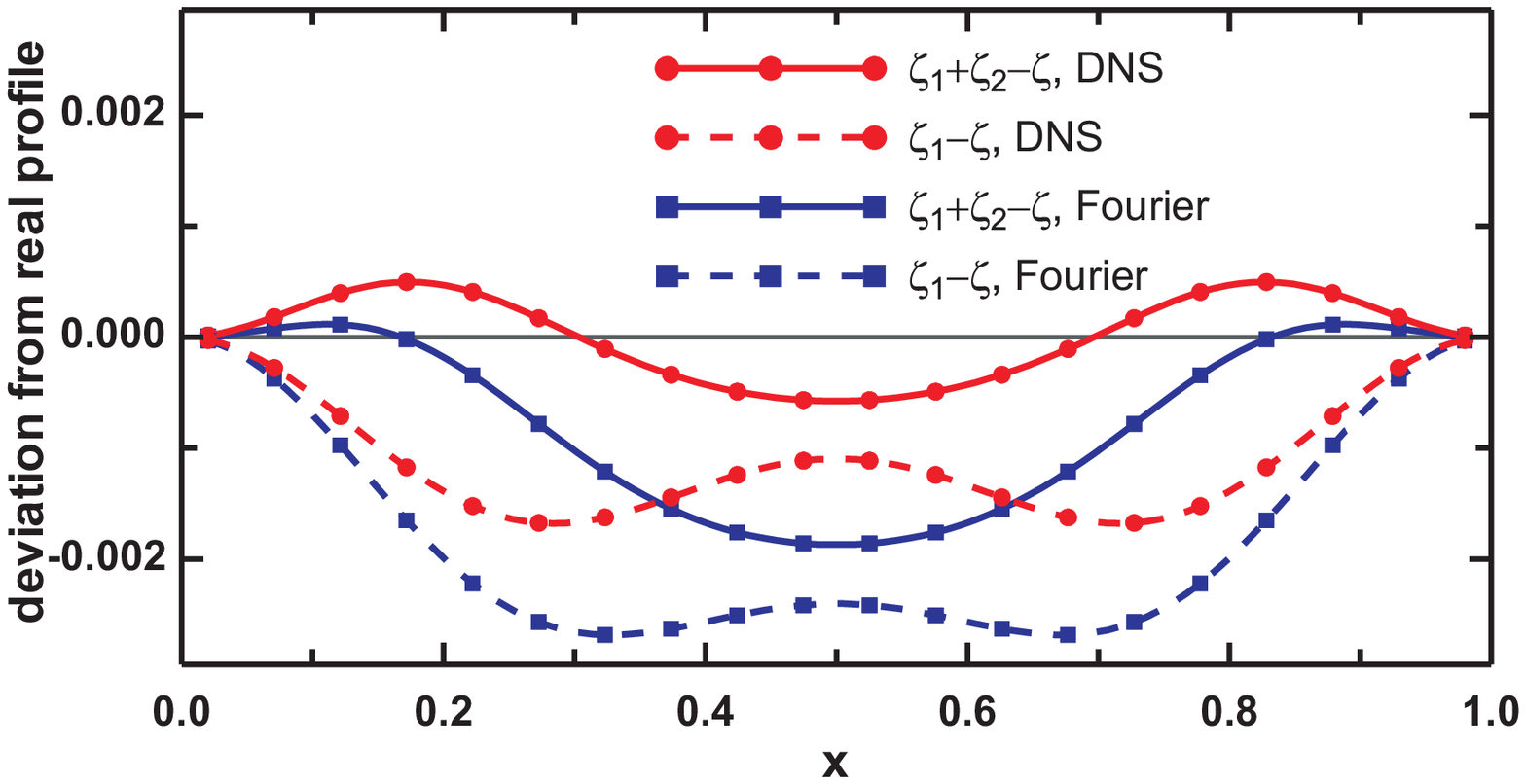}
  }

\caption{
(a):~The reconstructed profile~(\ref{eq116}) of a ferromagnetic layer (red dashed line) compared to the original profile (black solid line) for the parameter values $\zeta_0=0.1$, $a=0.01$, $k=2\pi$, $\mu_1/\mu_2=100$. (b):~Inaccuracy (the deviation from the original profile) of the profiles reconstructed from the solution with direct numerical simulation (DNS) and the linear-in-defect solution in Fourier space.
 }
\label{2dtesting}
\end{figure}

Linear in $(\zeta-\zeta_0)$ solution to the problem can be found analytically in Fourier space. Non-diverging for $z\to\infty$ solution to the problem reads
\begin{equation}
 \Phi(x,z)=\left\{
\begin{aligned}
-\mathcal{H}x+A\frac{\cosh{k\left|z\right|}}{\cosh{k\zeta_0}}\sin{kx},&\quad \left|z\right|<\zeta\,;\\[5pt]
-\mathcal{H}x+Ae^{-k(\left|z\right|-\zeta_0)}\sin{kx},&\quad \left|z\right|>\zeta\,.
\end{aligned}
\right.
\label{eq201}
\end{equation}
Here
\[
A=\frac{(\mu_1-\mu_2)\mathcal{H}a}{\mu_1\tanh{k\zeta_0}+\mu_2e^{-k\zeta_0}}\,.
\]
One can	differentiate the latter solution to find  the components of the magnetic field at height $h$;
\begin{equation}
\begin{aligned}
 H_x|_{z=h}&=\mathcal{H}-kA\,e^{-k(h-\zeta_0)}\cos{kx}\,,\\[5pt]
 H_z|_{z=h}&=kA\,e^{-k(h-\zeta_0)}\sin{kx}\,.
\end{aligned}
\label{eq202}
\end{equation}

In Figure~\ref{2dtesting}, with synthetic data from direct numerical simulation, one can see the surface profile can be well reconstructed with Eq.\,(\ref{eq116}). The accuracy of analytical solution (\ref{eq201}) and the role of correction $\zeta_2$ (compare Eqs.\,(\ref{eq114}), (\ref{eq115}) and  (\ref{eq116})) can be judged from Figure~\ref{2dtesting}(b).

\subsection{3D layer: measurable $\vec{H}$-field}
For the case of sensors measuring the $\vec{H}$-field at certain elevation above the layer with a dense enough grid of measurements points (e.g., \cite{Miller-Sander-2008}), one can directly employ Eqs.\,(\ref{eq125}) and (\ref{eq126}) with $\vec{H}$-field derivatives approximated by finite differences. If the elevation height is of the same order of magnitude as the characteristic defect width, one needs first to calculate $\vec{H}$-field on the ``imaginary'' undamaged surface and then use this field for calculation of the layer thickness profile. For this calculation one can use the Taylor expansion of $\vec{H}$-field.  Indeed, all the $z$-derivatives of $\vec{H}$-field can be calculated from $x$- and $y$-derivatives of measured fields, because $\vec{H}$ is a gradient of a harmonic function (see Eq.\,(\ref{eq101}));
\begin{align}
\frac{\partial{H_z}}{\partial{z}}&=
 -\left( \frac{\partial{H_x}}{\partial{x}}
        +\frac{\partial{H_y}}{\partial{y}} \right)\,,
\nonumber\\[5pt]
\frac{\partial^2H_z}{\partial{z}^2}&=
 -\left( \frac{\partial^2}{\partial{x}^2}
        +\frac{\partial^2}{\partial{y}^2} \right)H_z\,,
\nonumber\\[5pt]
\frac{\partial^{n+2}H_z}{\partial{z}^{n+2}}&=
 -\left( \frac{\partial^2}{\partial{x}^2}
        +\frac{\partial^2}{\partial{y}^2} \right)
        \frac{\partial^nH_z}{\partial{z}^n}\,,\quad n=1,2,3,\dots\,,
\nonumber
\end{align}
\begin{align}
\frac{\partial{H_x}}{\partial{z}}&=\frac{\partial{H_z}}{\partial{x}}\,,
\nonumber\\[5pt]
\frac{\partial^{n+2}H_x}{\partial{z}^{n+2}}&=
 -\left( \frac{\partial^2}{\partial{x}^2}
        +\frac{\partial^2}{\partial{y}^2} \right)
        \frac{\partial^nH_x}{\partial{z}^n}\,,\quad n=0,1,2,\dots\,,
\nonumber
\end{align}
\begin{align}
\frac{\partial{H_y}}{\partial{z}}&=\frac{\partial{H_z}}{\partial{y}}\,,
\nonumber\\[5pt]
\frac{\partial^{n+2}H_y}{\partial{z}^{n+2}}&=
 -\left( \frac{\partial^2}{\partial{x}^2}
        +\frac{\partial^2}{\partial{y}^2} \right)
        \frac{\partial^nH_y}{\partial{z}^n}\,,\quad n=0,1,2,\dots\,.
\nonumber
\end{align}

\subsection{3D layer: measurable $H_z$ or $(dH_z/dx)$}
When only one component of the magnetic field is measured, one has to make a more substantial use of the harmonic property of the field. Indeed, a harmonic function within $0\le x\le L_x$ and $0\le y\le L_y$, bounded at $z\to+\infty$, can be represented in the basis of exponentials;
\begin{equation}
\Phi(x,y,z)=\sum_{m,\,n}\phi_{mn}
 e^{i\frac{2\pi m}{L_x}x}e^{i\frac{2\pi n}{L_y}y}
 e^{-\sqrt{\left(\frac{2\pi m}{L_x}\right)^2+\left(\frac{2\pi n}{L_y}\right)^2}(z-h)}.
\label{eq203}
\end{equation}
Amplitudes $\phi_{mn}$ can be evaluated from Fourier decomposition of the measured component of the $H$-field (or its $x$-derivative)
\begin{equation}
H_z|_{z=h}=\sum_{m,\,n}\phi_{mn}
 \sqrt{\left(\frac{2\pi m}{L_x}\right)^2+\left(\frac{2\pi n}{L_y}\right)^2}
 e^{i\frac{2\pi m}{L_x}x}e^{i\frac{2\pi n}{L_y}y}.
\label{eq204}
\end{equation}
With $\phi_{mn}$ known, one can calculate derivatives of $\Phi(x,y,z)$ at any $z$, which are components of the $\vec{H}$-field, and employ Eqs.\,(\ref{eq125}) and (\ref{eq126}).

\section{Conclusion}
We have developed a technique for the analytical calculation of ferromagnetic casing thickness profiles from measurements of the magnetic field above the layer when a homogeneous external magnetic field is applied, i.e., for the magnetic flux leakage (MFL) method for inspection of wellbore casing integrity. The analytical results have been derived within the framework of the large-scale approximation of defects, the widths of which is large compared to their depth and layer thickness; this approximation is generally relevant for corrosion damage (with the exception of corrosive damage of welds). The technique has been shown to be applicable for a nonlinear magnetisation law and without hysteresis within the working range of $H$-field strength. The latter restriction potentially diminishes the applicability of the result, but MFL tools are designed to saturate the casing to minimize the impact of hysteresis on the analysis. The applicability of the analytical results has been validated with the results of direct numerical simulation.


\end{document}